\documentclass[12pt,halfline,a4paper]{ouparticle}

\usepackage{graphics} 
\usepackage{graphicx} 
\usepackage{amsmath} 
\usepackage{amssymb}  
\usepackage[utf8]{inputenc}
\usepackage[english]{babel}
\usepackage[numbers]{natbib}
\usepackage{appendix}
\usepackage{arydshln}
\usepackage[labelfont=bf]{caption}
\usepackage[section]{placeins}
\usepackage{multirow}
\usepackage{booktabs}

\begin{document}
	
	\title{\textbf{\break {Parameter Estimation in Mean Reversion Processes with Periodic Functional Tendency}}}
	
	\author{%
		\name{Juan Pablo Pérez Monsalve}
		\address{Universidad EAFIT, Department of Finance\\
			Medellín, Colombia}
		\email{jperezm9@eafit.edu.co}
		\and
		\name{Freddy H. Marín Sanchez}
		\address{Universidad EAFIT, Department of Mathematical Sciences\\
			Medellín, Colombia}
		\email{fmarinsa@eafit.edu.co}}
	

	\abstract{This paper describes the procedure to estimate the parameters in mean reversion processes with functional tendency defined by a periodic continuous deterministic function, expressed as a series of truncated Fourier. Two phases of estimation are defined, in the first phase through Gaussian techniques using the Euler-Maruyama discretization, we obtain the maximum likelihood function, that will allow us to find estimators of the external parameters and an estimation of the expected value of the process. In the second phase, a reestimate of the periodic functional tendency with it's parameters of phase and amplitude is carried out, this will allow, improve the initial estimation. Some experimental result using simulated data sets are graphically illustrated.} 
	
	\date{}
	
	
	\maketitle
	
	
\section{Introduction}

Stochastic models have applications in many disciplines including engineering, economics and finance, physics, biology and medicine, allowing to identify the dynamics of natural phenomena, physical phenomena, asset prices and population growth, among others. Particularly, the mean reversion processes occupy an important place within the stochastic models, whose application stands out in fields such as the energy markets \cite{Pilipovic2007}, commodity prices, bonds and interest rates (See, e.g., \cite{MarinSanchez2013} and references therein.).

The mean reversion processes tend to grow (decrease) after reaching a minimum (maximum) \cite{Exley2004}, therefore, they oscillate around some level of equilibrium. Thus, when the stochastic term adds volatility to the process causing it to move away from the equilibrium level, the deterministic term that acts as a trend will allow the process to return to that equilibrium level \cite{Pilipovic2007}. 

Although the trend term can be defined from a constant parameter \cite{Sanchez2016}, is more useful to define the trend as a deterministic function or as another stochastic process \cite{Tifenbach2000}. In this sense, the deterministic function allows to identify the essence of the trend of the process, tendency that is generally unknown and that in a great part of the processes follows a periodic or almost periodic dynamics  or that is affected by cycles. This models have been studied particularly in the field of finance to describe the dynamics of commodities and underlying assets in derivatives. Thus, the model proposed by \cite{Pilipovic2007} is used to capture the dynamics of the price of electric energy, while \cite{Alaton2002} for their part, model the temperature through a stochastic differential equation with functional tendency to find a price model of a derivative on climate.

The approaches used for the specification of stochastic models of mean reversion processes are based on the a priori knowledge of the process or on statistical methods that allow the parametric representation of such process based on the available historical information. Since the model specification is made in continuous time and the process trajectories are defined in discrete time, a discretization of the model is necessary, for which the Euler-Maruyama scheme is usually used. From this scheme, a stochastic model in discrete time is obtained, and on this are realized statistical inferences such as the estimation of parameters.

There are different procedures for the estimation of parameters such as distributional moments, Kernel method, and ordinary least squares and maximum likelihood as discussed in \cite{Sanchez2016,MarinSanchez2013}. In particular, this last method has received considerable attention during the last decades, since researchers in empirical finance have used these models in applications of importance to the financial industry \cite{Phillips2009}, where stand out researches made by \cite{Durham2002,Fergusson2015,Kleppe2014,Phillips2009a,Tang2009}. Thus, this method is based on the construction of a likelihood function derived from the transition probability density of discretely sampled data, then closed-form sequences are used to approximate the transition density which is equivalent to an approximation to the likelihood function. Within the approximation mechanisms, exist those based on Hermite polynomial expansions \cite{AitSahalia2002}, while the others are based on the saddlepoint approximation \cite{Phillips1978,A2006}.

The objective of this paper is to estimate the parameters of one-factor mean reversion stochastic models where the functional trend follows a periodic behavior defined by a series of truncated Fourier, for it, the estimation of maximum likelihood is used from the discretization of the model. In this sense, both the periodic trend function and the parameters are estimated. To estimate the trend function in a first moment, smoothing techniques and numerical derivatives are used, while in the second moment the Fourier analysis is executed, in particular the Discrete Fourier Transform (DFT) which gives a functional representation of the trend. The parameters are estimated in two moments using the maximum likelihood technique from the properties of the discretized error normality.

This paper is organized as follows. In section \ref{Fanalysis.1} we show the Fourier analysis that serves as the basis for the representation of the periodic functional trend. A detailed description of the mean reversion processes with periodic functional trend is shown in section \ref{meanreversion.1}. Section \ref{estimationm.1} develops the proposed model by describing the two estimation phases for the parameters and the periodic functional trend. In section \ref{results.1} we show some numerical examples of the estimation for the one particular case, other numerical examples are show in the appendix. Finally the conclusions are exhibit in section \ref{conclusions.1}.

\section{Fourier Analysis}\label{Fanalysis.1}

A signal is a representation of the information, whose manipulation is an important activity in the field of science and engineering, since it can be easily treated if is represented as a linear combination of simple and mathematically well defined signals \cite{Sundararajan2001}. It is at this point where the Fourier theory is important.

Since the work titled \textit{Theorie Analytique de la Chaleur}, Fourier has set a precedent in the mathematical field, given that the series and the transformations that bear his name, have contributed in the theory of the partial differential equations, the harmonic analysis, the theory of the representation, the theory of the numbers and the geometry \cite{Wong2011}, with important implications in the field applied in particular signal processing \cite{Madisetti1998}. In this context, the Fourier's great contribution was to show that (at least mathematically) any phenomenon defined by a restricted motion (periodic) could be expressed as the combined output of a number of sinusoidal generators \cite{Prandoni2008}. 

The Fourier series is the representation in the frequency domain of a continuous time periodic signal in terms of an infinite set of  harmlessly related sinusoids \cite{Sundararajan2001}. According to \cite{Madisetti1998} the conditions under which a periodic signal $s(t)$ can be expanded in a Fourier series are known as the Dirichet conditions,  in this sense in each period, $s(t)$ must have a finite number of discontinuities, a finite number of maxims and minims, and that $s(t)$ satisfies the absolute convergence criterion defined by $\int_{-T/2}^{T/2}\left|s(t)\right|dt<\infty$. Thus, if at a continuous time the signal $s(t)$ is periodic with a period $T$ the complex representation of the Fourier series of $s(t)$ is given by,

\begin{equation}
\label{compleja.1}
s(t)=\sum_{n=-\infty}^{\infty}a_{n}e^{jnw_{0}t}
\end{equation}

With $w_{0}=\frac{2\pi}{T}$ and where $a_{n}$ are the complex Fourier coefficients given by,

\begin{equation*}
a_{n}=\frac{1}{T}\int_{-T/2}^{T/2}s(t)e^{-jnw_{0}t}dt
\end{equation*}

For each value of $t$ where $s(t)$ is continuous, the right side of (\ref{compleja.1}) converges to $s(t)$. The complex representation of the Fourier series in (\ref{compleja.1}), can be manipulated to obtain a trigonometric expression that contains the terms $\sin(w_{0}t)$ and $\cos(w_{0}t)$, thus, the trigonometric form of the Fourier series for a signal $s(t)$ is given by, 

\begin{equation*}
s(t)=\sum_{n=0}^{\infty}b_{n}\cos(nw_{0}t)+\sum_{n=1}^{\infty}c_{n}\sin(nw_{0}t)
\end{equation*}

Where, $w_{0}=\frac{2\pi}{T}$, $b_{n}$ and $c_{n}$ are Fourier coefficients such that,

\begin{equation*}
\begin{split}
b_{0}&=\frac{1}{T}\int_{-T/2}^{T/2}s(t)dt\\
b_{n}&=\frac{2}{T}\int_{-T/2}^{T/2}s(t)\cos(nw_{0}t)dt \hspace{0.4cm} n=1,2,\cdot\cdot\cdot,\\
c_{n}&=\frac{2}{T}\int_{-T/2}^{T/2}s(t)\sin(nw_{0}t)dt  \hspace{0.4cm} n=1,2,\cdot\cdot\cdot,
\end{split}
\end{equation*}

According to \cite{Madisetti1998} the Fourier series is a classical Fourier method in which, the analysis is performed in continuous time, that is, $s(t)$ is defined for all values of $t$ in the continuum $-\infty<t<\infty$. More recent developments include the so-called Discrete Time Fourier Transform (DTFT) and Discrete Fourier Transform (DFT), which are extensions of Fourier concepts that apply to discrete-time signals, with the particularity that such signal is defined only for values of $n$ where $n$ is an integer in the range $-\infty<n<\infty$.

In this sense, the Fourier transform in its discrete version is appropriate for the harmonic analysis of discrete data such as those obtained from experimental measurements or by sampling a function in a finite set of points \cite{Hanna2012}. Thus, the DFT of a signal is an alternative representation of the data in the signal, therefore, while a signal lives in the time domain, its Fourier representation lives in the domain of the frequency \cite{Prandoni2008}. 

According to \cite{Bremaud2013}, in practical terms we have an infinite vector of samples $s=(s_{0},\cdot\cdot\cdot,s_{N-1})$ for a signal $s(t)$, where $s_{n}=s(n\Delta)$, then the sum of Fourier of this vector evaluated in the pulses $w_{k}=\frac{2k\pi}{N}$ corresponds to the Discrete Fourier Transform. In this sense, the DFT of $s=(s_{0},\cdot\cdot\cdot,s_{N-1})$, is the vector $S=(S_{0},\cdot\cdot\cdot,S_{N-1})$, where,

\begin{equation}
\label{fou1.1}
\begin{split}
S_{k}&=\sum_{n=0}^{N-1}s_{n}e^{\frac{-i(2\pi kn)}{N}} \hspace{0.4cm} \text{with} \hspace{0.4cm} k=0,1,\cdot\cdot\cdot, N-1\\
s_{n}&=\frac{1}{N}\sum_{k=0}^{N-1}S_{k}e^{\frac{i(2\pi kn)}{N}} \hspace{0.4cm} \text{with} \hspace{0.4cm} n=0,1,\cdot\cdot\cdot, N-1
\end{split}
\end{equation}

Regardless of whether $S_{n}$ has a finite length or a periodic sequence, the DFT treats the $N$ samples of $s_{n}$ as though they are one period of a periodic sequence \cite{Madisetti1998}, therefore the processing of the signal made with the DFT will inherit the consequences of this assumed periodicity. Note that the DFT is an approximation to the Fourier transform whose quality depends on the parameters $N$ and $\Delta$.

In this order, following \cite{Prandoni2008} with the DFT, we obtain the decomposition of a finite-length signal $s(t)$ into a set of $N$ sinusoidal components, where the magnitude and initial phase of each oscillator are given by the coefficients $S_{k}$. Thus, the DFT takes a series of $N$ complex sinusoidal generators, sets the frequency of the k-th generator to $\left(\frac{2\pi k}{N}\right)$, sets the amplitude of the k-th generator to $\left|S_{k}\right|$, that is, the magnitude of the k-th coefficient DFT, sets the phase of the k-th generator to $\measuredangle S_{k}$ i.e. the phase of the k-th coefficient DFT, and finally starts the generators at the same time and adds their outputs. 

\section{Mean Reversion Processes with Periodic Functional Tendency}\label{meanreversion.1}

Mean reversion processes with periodic functional tendency can be written as a linear stochastic differential equation of form,

\begin{equation}
\label{ede.1}
dX_{t}=\alpha (\mu(t)-X_{t})d_{t}+\sigma X_{t}^{\gamma}dB_{t}
\end{equation}

With the initial condition $X_{0}=x$, where $\alpha>0$, $\sigma>0$, $\gamma=\left \{0,\frac{1}{2},1\right \}$ are constants, $\mu(t)$ is a continuous deterministic function of values in $\mathbb{R}$ defined by the series of Fourier $\mu(t)=\sum_{k=0}^{n}a_{k}\cos(2\pi tk+\phi_k)$ with $n=0,1,2,\cdot\cdot\cdot$, and $\left \{B_{t}\right \}_{t\geq 0}$ is a One-dimensional Standard Brownian Motion defined in a probability space $(\Omega,\mathcal{F},\mathbb{P})$. Appendix \ref{Anexo1.1} shows that the Lipschitz and growth conditions are satisfied for both cases $\gamma=0$ and $\gamma=1$, for $\gamma=\frac{1}{2}$ the procedure is similar.    

The parameter $\alpha$ is defined as the rate of reversion, while $\mu(t)$ is the mean reversion level, $\sigma$ is the parameter associated with volatility and $\gamma$ determines the sensitivity of the variance to the level of $X_{t}$. It should be noted that when $\gamma=0$ this says that the process of mean reversion with functional tendency has additive noise, when $\gamma=1$ the process has proportional noise and when $\gamma=\frac{1}{2}$ a non-linear stochastic equation is obtained.  

The model (\ref{ede.1}) is a generalization of the CKLS model, proposed by \cite{Chan1992}, so the mean reversion level is a deterministic function that captures the trend of process. $\mu(t)$ is constituted as an attractor at each point $t$ in the sense that, when $X_{t}>\mu(t)$ the trend term $\alpha(\mu(t)-X_{t})<0$ and therefore $X_{t}$ decreases, in the case in which $X_{t}<\mu(t)$ it means that $X_{t}$ grows. 

Equation (\ref{ede.1}) can be written in its integral form as,

\begin{equation}
X_{t}-X_{0}=\alpha\int_{0}^{t}(\mu(s)-X_{s})ds+\sigma\int_{0}^{t}X_{t}^{\gamma}dB_{s}
\end{equation}

Taking the expected value, $E[X_{t}]-E[X_{0}]=\alpha\int_{0}^{t}(\mu(s)-E[X_{s}])ds$, and thus we obtain the differential equation, 

\begin{equation}
\stackrel{.}{m}(t)=\alpha(\mu(t)-m(t)) \hspace{0.4cm} m(t)=E[X_{t}]
\end{equation}

The solution of this equation is, $m(t)=m(0)e^{-\alpha t}+\alpha e^{-\alpha t}\int_{0}^{t}\mu(s)e^{\alpha s}ds$, with $\mu(t)=\sum_{k=0}^{n}a_{k}\cos(2\pi tk+\phi_k)$, then, 

\begin{equation*}
m(t)=m_{0}e^{-\alpha t}+\alpha e^{-\alpha t}f(t) \hspace{0.4cm} \text{with},
\end{equation*}

\begin{equation}
f(t)=\sum_{k=0}^{n}\frac{a_{k}\left(e^{\alpha t}(\alpha\cos(2\pi kt+\phi_{k})+2\pi k\sin(2\pi kt+\phi_{k}))-\alpha\cos(\phi_{k})-2\pi k\sin(\phi_{k})\right)}{\alpha^{2}+(2\pi k)^{2}}
\end{equation}

\begin{figure}[h]
	\centering
	\includegraphics[width=0.89\linewidth]{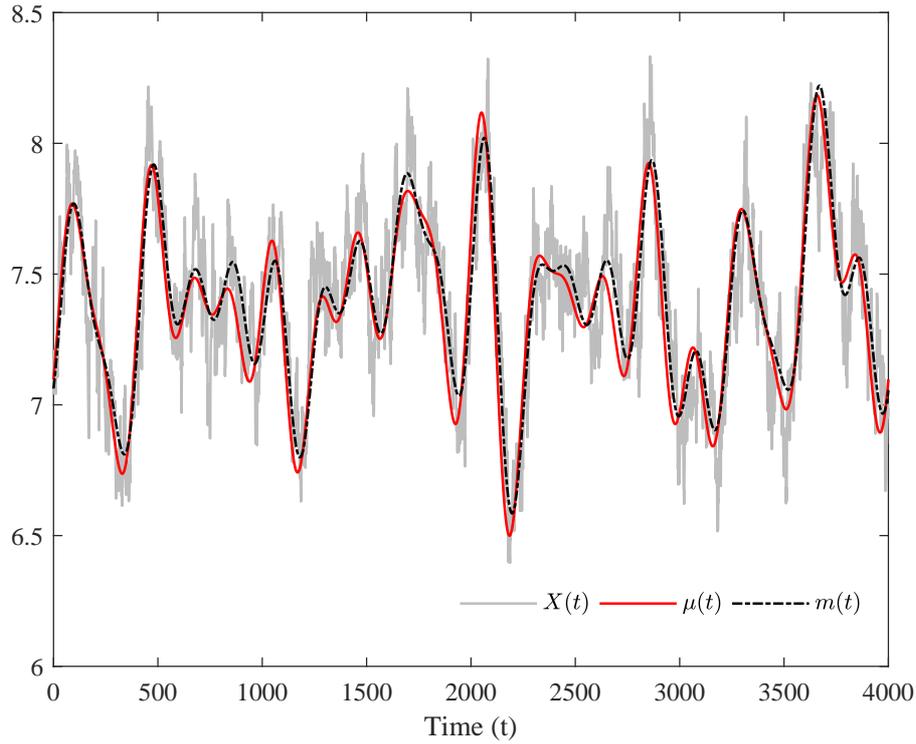}
	\caption{\textbf{Dynamic behavior of $X(t)$, $\mu(t)$ and $m(t)$}}
	\medskip
	\begin{minipage}{0.64\textwidth} 
		{\footnotesize{The parameters with which $X(t)$ and $\mu(t)$ are calculated are defined in Table \ref{tab:tabla1-1}, $m(t)$ is the expected value of $X(t)$. With $\Delta t=\frac{1}{250}$ and a total of 4000 observations.\par}}
	\end{minipage}
	\label{fig:1}
\end{figure}

The Figure \ref{fig:1} shows the dynamic behavior of $\mu(t)$, $m(t)$ and one path of $X_{t}$ for specific values of the parameters which are observed in Table \ref{tab:tabla1-1}. This figure shows how the expected value of the process is close to the trend level $\mu(t)$, for $\Delta t=\frac{1}{250}$ and a total of 4000 observations. 

\begin{table}[htbp]
	\centering
	\caption{\textbf{Base Parameters}}
	\resizebox{\textwidth}{!}{
	\begin{tabular}{ccccccccccccccccccccc}
		\toprule
		\toprule
		\multicolumn{21}{c}{\textbf{Parameters}} \\
		\cmidrule{1-21}
		\multicolumn{21}{c}{\textbf{$\alpha=20 \hspace{0.4cm} \sigma=1.1 \hspace{0.4cm} \gamma=0$}} \\
		\hdashline
		$k$    & & 0 &    & 2 &    & 4 &    & 9  &   & 10 &   & 12 &   & 13  &  & 15  &  & 16 &   & 20 \\
		$a_{k}$ &   & 7.3728 && 0.0786 && 0.1664 && 0.1576 && 0.2074 && 0.1376 && 0.1380 && 0.1626 && 0.0964 && 0.1756 \\
		$\phi_{k}$  && 0     && 0.6331 && 2.0853 && -2.1316 && -1.4149 && -1.0862 && 2.6551 && 2.0512 && -1.8092 && -1.8587 \\
		\bottomrule
		\bottomrule
		\multicolumn{21}{p{21cm}}{\footnotesize{The internal input parameters are randomly defined up to a $k=20$. The amplitude parameters (with the exception of $a_{0}$) were chosen randomly for a range between 0.006 and 0.22, the $k$ indicator was randomly chosen between 0 and 20 and the phase angle $\phi_{k}$ was chosen randomly between -2.9 and 2.9.}}
	\end{tabular}}
	\label{tab:tabla1-1}
\end{table}

\section{Gaussian Estimation Method}\label{estimationm.1}

\subsection{First Phase of Estimation}\label{firstphase.1}

According  to the methodology used in \cite{MarinSanchez2013} and \cite{Sanchez2016}, is possible to obtain closed formulas for the parameter estimators from the discrete observations of a path of process. Consider the differential equation (\ref{ede.1}) with $X_{0}=x$ as the initial value, $\alpha(\mu(t)-X_{t})$ as the trend function, $\sigma X_{t}^{\gamma}$ as the diffusion function and $\Theta=[\alpha,\gamma,\sigma]$ as a unknown parameter vector, where the conditions of existence and uniqueness are guaranteed (see appendix \ref{Anexo1.1}), thus the solution of $X_{t}$ exists and $\mu(t)$ can be defined in the form,

\begin{equation}
\label{mudef.1}
\mu(t)=m(t)+\frac{\stackrel{.}{m}(t)}{\alpha}
\end{equation}

Replacing in (\ref{ede.1}), 

\begin{equation}
\label{eseuler.1}
dX_{t}=\alpha \left(m(t)+\frac{\stackrel{.}{m}(t)}{\alpha}-X_{t}\right)d_{t}+\sigma X_{t}^{\gamma}dB_{t}
\end{equation}

Using the numerical scheme of Euler-Maruyama to equation (\ref{eseuler.1}) and defining a new variable $Y_{t}$,

\begin{equation*}
Y_{t}=\frac{X_{t}-X_{t-1}-[\alpha(m_{t-1}-X_{t-1})+\stackrel{.}{m}(t)]\Delta}{X_{t-1}^{\gamma}}=\epsilon_{t}  \hspace{0.15cm} ; \hspace{0.4cm} \epsilon_{t} \sim N(0,\sigma^{2}\Delta) 
\end{equation*}

\begin{figure}[h]
	\centering
	\includegraphics[width=0.89\linewidth]{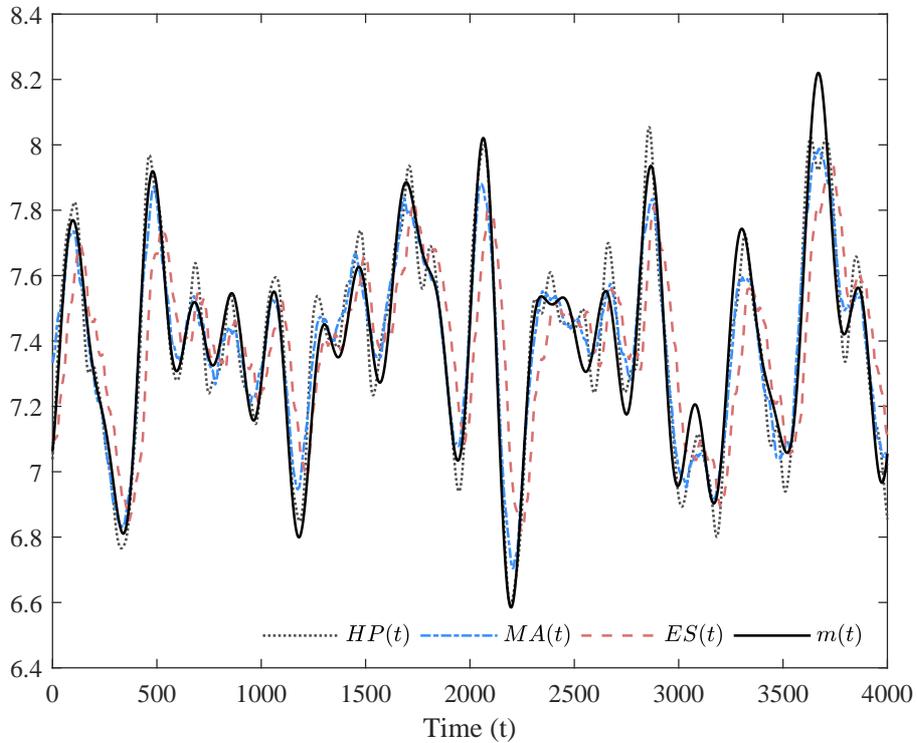}
	\caption{\textbf{Dynamic behavior of $HP(t)$, $MA(t)$, $ES(t)$ and $m(t)$}}
	\medskip
	\begin{minipage}{0.775\textwidth} 
		{\footnotesize{$HP(t)$ is the Hodrick-Prescott filter where the smoothing parameter is 40000, $MA(t)$ is a Moving Average calculated over a sliding window with 100 observations of length, $ES(t)$ is an Exponential Smoothing where period length is 1000 and  $m(t)$ is the expected value of $X(t)$. With $\Delta t=\frac{1}{250}$ and a total of 4000 observations.\par}}
	\end{minipage}
	\label{fig:2}
\end{figure}

Since the variable $Y_{t}$ depends on the observations of $X_t$, the unobserved variable $m(t)$ and its derivative $\stackrel{.}{m}(t) $, is necessary to estimate the expected value $m(t)$ from the observations of the process, considering that the path of the sample summarizes all the information that is known of that process. Thus, the estimate of $m(t)$ can be obtained using different smoothing techniques which may include filters and other smoothing as moving averages, as described in \cite{Sanchez2016}. Figure \ref{fig:2} shows the dynamics of $m(t)$ and the dynamics of the Hodrick-Prescott filter (HP), a moving average (MA) and an exponential smoothing (ES) applied on $X(t)$.

Once the estimate of $m(t)$ is obtained from the observations of $X_{t}$, we proceed to obtain $\stackrel{.}{m}(t)$, using numerical derivation techniques following Taylor's theorem, such as the three-point rule, or the five-point rule.

Given the estimation of $m(t)$ and $\stackrel{.}{m}(t)$, we can construct one realization of $Y_{t}$ and proceed to define its normal density function,

\begin{equation*}
\begin{split}
f(Y_{i}:\theta)&=\left(\frac{1}{2\pi\sigma^{2}\Delta}\right)^{1/2}\cdot\\
&\exp\left[\frac{-1}{2\sigma^{2}\Delta}\left(\frac{X_{i}-X_{i-1}-[\alpha(m_{i-1}-X_{i-1})+\stackrel{.}{m}(i-1)]\Delta}{X_{i-1}^{\gamma}}\right)^{2}\right]
\end{split}
\end{equation*} 

Where the joint density function is given by,

\begin{equation*}
f(Y_{1},Y_{2},\cdot\cdot\cdot,Y_{T})=\prod_{k=1}^{T}f(Y_{k}:\theta)
\end{equation*} 

And the maximum likelihood function results in,

\begin{equation*}
\begin{split}
L(\theta|\left\{Y_{t}\right\})&=\left(\frac{1}{2\pi\sigma^{2}\Delta}\right)^{T/2}\cdot\\
&\exp\left[\frac{-1}{2\sigma^{2}\Delta}\sum_{i=1}^{T}\left(\frac{X_{i}-X_{i-1}-[\alpha(m_{i-1}-X_{i-1})+\stackrel{.}{m}(i-1)]\Delta}{X_{i-1}^{\gamma}}\right)^{2}\right]
\end{split}
\end{equation*}

In this sense the problem of maximizing the likelihood function is given by,

\newcommand{\Arg}{\mathop{\mathrm{arg}}\limits} 

\begin{equation*}
\frac{\partial\log(L)}{\partial\theta}=\stackrel{\to }{0}; \hspace{0.4cm} \hat{\theta}=\Arg_{\theta} \max(\log(L)) 
\end{equation*}  

Assuming $\gamma$ as known the estimation for $\alpha$ and $\sigma$ is given by,

\begin{equation*}
\begin{split}
\hat{\alpha}&=\frac{\sum_{i=1}^{T}\left((X_{i}-X_{i-1}-\stackrel{.}{m}(i-1)\Delta)(m_{i-1}-X_{i-1})/X_{i-1}^{2\gamma}\right)}{\sum_{i=1}^{T}\left[(m_{i-1}-X_{i-1})/X_{i-1}^{\gamma})\right]^{2}\Delta}\\
\\
\hat{\sigma}&=\sqrt{\frac{1}{T\Delta}\sum_{i=1}^{T}\left(\frac{X_{i}-X_{i-1}-[\hat{\alpha}(m_{i-1}-X_{i-1})+\stackrel{.}{m}(i-1)]\Delta}{X_{i-1}^{\gamma}}\right)^{2}}
\end{split}
\end{equation*}

In this sense, considering $\hat{\alpha}$ and (\ref{mudef.1}), a first estimate of $\mu(t)$ can be obtained.

\subsection{Second Phase of Estimation}

Once $\hat{\mu}(t)$ is found, an estimate of $\mu(t)$ that is more precise can be obtained. To do this, we proceed to approximate $\hat{\mu}(t)$ through the Fourier analysis, so we take the observations of $\hat{\mu}(t)$ to obtain $\hat{\hat{\mu}}(t)$.

In this sense, as defined in section \ref{Fanalysis.1}, by having a sample vector $\hat{\mu}=(\hat{\mu}_{0},\cdot\cdot\cdot,\hat{\mu}_{N-1})$ for a signal $\hat{\mu}(t)$, where $\hat{\mu}_{n}=\hat{\mu}(n\Delta)$, is possible to obtain a vector of complex numbers $\hat{M}=(\hat{M}_{0},\cdot\cdot\cdot,\hat{M}_{N-1})$ through the DFT, where,

\begin{equation}
\label{dft1.1}
\begin{split}
\hat{M}_{k}&=\sum_{n=0}^{N-1}\hat{\mu}_{n}e^{\frac{-i(2\pi kn)}{N}} \hspace{0.4cm} \text{with} \hspace{0.4cm} k=0,1,\cdot\cdot\cdot, N-1\\
\hat{\mu}_{n}&=\frac{1}{N}\sum_{k=0}^{N-1}\hat{M}_{k}e^{\frac{i(2\pi kn)}{N}} \hspace{0.4cm} \text{with} \hspace{0.4cm} n=0,1,\cdot\cdot\cdot, N-1
\end{split}
\end{equation}

This equation indicates that the DFT is a function of the discrete variable $k$. In general the values given by the DFT are complex, therefore $\hat{M}_{k}$ can be expressed in rectangular form as shown in \cite{Oppenheim1975}, such that,

\begin{equation*}
\hat{M}_{k}=R_{k}+iI_{k}
\end{equation*} 

Where $R_{k}$ corresponds to the real part of $\hat{M}_{k}$, while $I_{k}$ corresponds to the imaginary part of $\hat{M}_{k}$. Thus, using Euler's formula, the equation (\ref{dft1.1}) becomes,

\begin{equation*}
\hat{M}_{k}=\sum_{n=0}^{N-1}\hat{\mu}_{n}\left(\cos\left(-\frac{2\pi nk}{N}\right)+i\sin\left(-\frac{2\pi nk}{N}\right)\right)
\end{equation*}

Such that,

\begin{equation*}
\begin{split}
R(\hat{M}_{k})&=\sum_{n=0}^{N-1}\hat{\mu}_{n}\cdot\cos\left(-\frac{2\pi nk}{N}\right)\\
I(\hat{M}_{k})&=\sum_{n=0}^{N-1}\hat{\mu}_{n}\cdot\sin\left(-\frac{2\pi nk}{N}\right)
\end{split}
\end{equation*}

Since the discrete Fourier coefficients are cyclic with period $N$, developing the complex exponential of (\ref{dft1.1}) and regrouping terms can represent this expression in its first trigonometric form, such that,

\begin{equation}
\label{representation.1}
\hat{\hat{\mu}}_{n}=\sum_{k=o}^{L}a_{k}\cos\left[\frac{2\pi kn}{N}+\phi_{k}\right]
\end{equation}

Where $L=\frac{N}{2}$ for $N$ even and $L=\frac{N-1}{2}$ for $N$ odd. Then, 

\begin{equation*}
\begin{split}
a_{k}&=\left|\hat{M}_{k}\right|=\sqrt{R(\hat{M}_{k})^{2}+I(\hat{M}_{k})^{2}}\\
\phi_{k}&=\arg(\hat{M}_{k})=tan^{-1}\left(\frac{I(\hat{M}_{k})}{R(\hat{M}_{k})}\right)
\end{split}
\end{equation*} 

With $1$ and $\frac{1}{N}$ as the normalization factors that multiply to the DFT and its inverse respectively.

Under this context, the Fourier analysis is used to obtain a realization of $\hat{\mu}(t)$ which is equivalent to obtaining $\hat{\hat{\mu}}(t)$. Thus, from the Fourier analysis the parameters $a_{k}$ and $\phi_{k}$ are obtained, which allow to represent the trend level $\mu(t)$ with the smallest error. Once the best estimate of $\mu(t)$ is found, i.e. $\hat{\hat{\mu}}(t)$, we proceed to re-estimate the parameters $\alpha$ and $\sigma$ applying a mechanism similar to that used to obtain its first estimate in the section (\ref{firstphase.1}), defining a new variable,

\begin{equation*}
\tilde{Y_{t}}=\frac{X_{t}-X_{t-1}-\alpha(\mu_{t-1}-X_{t-1})\Delta}{X_{t-1}^{\gamma}}
\end{equation*} 

Then the re-estimation of $\alpha$ and $\sigma$ is given by,

\begin{equation*}
\begin{split}
\hat{\hat{\alpha}}&=\frac{\sum_{i=1}^{T}\left((X_{i}-X_{i-1})(\hat{\hat{\mu}}_{i-1}-X_{i-1})/X_{i-1}^{2\gamma}\right)}{\sum_{i=1}^{T}\left[(\hat{\hat{\mu}}_{i-1}-X_{i-1})/X_{i-1}^{\gamma})\right]^{2}\Delta}\\
\\
\hat{\hat{\sigma}}&=\sqrt{\frac{1}{T\Delta}\sum_{i=1}^{T}\left(\frac{X_{i}-X_{i-1}-\hat{\hat{\alpha}}(\hat{\hat{\mu}}_{i-1}-X_{i-1})\Delta}{X_{i-1}^{\gamma}}\right)^{2}}
\end{split}
\end{equation*}

\section{Results of Estimation}\label{results.1}

The objective of this section is to show the performance of the first and second estimation phase defined in the previous section. To achieve this, we will simulate paths of a process defined by equation (\ref{ede.1}), with a known parameter vector $\Theta=[\alpha,\gamma,\sigma]$ and a deterministic function $\mu(t)$, then we make the estimates with the established procedure, to see if these estimates are closer to the real parameters.

The level of trend $\mu(t)$ is simulated assuming a sinusoidal sum with 10 cosines where the parameters of amplitude, phase angle and indicator are chosen randomly for specific ranges, and with a parameter vector $\Theta$ known as shown in the Table \ref{tab:tabla1-1}, each path was generated for a total of 4000 observations with $\Delta t=\frac{1}{250}$. 

In this sense, from 1000 different simulations we proceed to make the estimates, thus, the expected value $\hat{m}(t)$ for each path is approximated by the Hodrick-Prescott filter with a smoothing parameter $\lambda=40000$, while $\hat{\stackrel{.}{m}}(t)$ is obtained through of a three-point numerical derivation rule. Once we have $\hat{m}(t)$ and $\hat{\stackrel{.}{m}}(t)$ the first estimation phase is executed, resulting in $\hat{\alpha}$, $\hat{\sigma}$ and $\hat{\mu}(t)$. Subsequently, the Fourier analysis is executed taking as signal the first estimate of $\mu(t)$, which results in $\hat{\hat{\mu}}(t)$, for this case the DFT is used assuming a sinusoidal sum with a total of 10 cosines, such that an expression as in (\ref{representation.1}) is obtained with a total of 10 amplitude parameters ($\hat{a}_{k}$), 10 phase angles ($\hat{\phi}_{k}$) and 10 indicators ($k$). This sinusoidal sum is chosen because with a number of 10 cosines the Fourier analysis captures sufficient information, thus, Table \ref{tab:tabla1.1-1} shows the RMS of $\hat{\hat{\mu}}(t)$ with L sinusoidal sums, with respect to $\hat{\hat{\mu}}(t)$ with L-1 sinusoidal sums, calculated from the average for each point in time of the paths of  $\hat{\hat{\mu}}(t)$ for each sinusoidal sum with 6, 7, 8, 9, 10, 11, 12 and 13 cosines. In this way, the RMS is reduced when more cosines are added, however, from 10 cosines onwards the RMS is not statistically different.

\begin{table}[htbp]
	\centering
	\caption{\textbf{RMS for L Sinusoidal Sums}}
	\resizebox{\textwidth}{!}{
	\begin{tabular}{ccccccccccccccc}
		\toprule
		\toprule
		\textbf{L-sum} & &7 &    & 8 &    & 9 &    & 10 &   & 11&    & 12  &&  13 \\
		\textbf{ RMS } & &0.001960&  & 0.001923  & &0.002037  & & 0.001307&  & 0.000004 &  & 0.000001 & & 0.000000  \\
		\bottomrule
		\bottomrule
		\multicolumn{15}{p{18cm}}{\footnotesize{1000 paths of $\hat{\hat{\mu}}(t)$ are calculated with Fourier analysis by each sinusoidal sum with cosines 6, 7, 8, 9, 10, 11, 12 and 13. Then, the RMS of $\hat{\hat{\mu}}(t)$ with L sinusoidal sums, with respect to $\hat{\hat{\mu}}(t)$ with L-1 sinusoidal sums is calculated, thus, the RMS of L-sum 7 is calculated as $\frac{\sum (\hat{\hat{\mu}}(t)_{6}-\hat{\hat{\mu}}(t)_{7})^{2}}{n}$ where $\hat{\hat{\mu}}(t)_{6}$ is the average for each point in time of the paths of $\hat{\hat{\mu}}(t)$ with 6 cosines in the sinusoidal sum.}}
	\end{tabular}}
	\label{tab:tabla1.1-1}
\end{table}

In Table \ref{tab:tabla2.1} we present the basic statistics (mean, median, mode and standard deviation) for the estimates obtained in the first and second phase, of external parameters $\alpha$ and $\sigma$. The results for $\alpha$ indicate that the average re-estimate is closer to the real value compared to the first estimate, in addition, the dispersion of the re-estimate is much lower than the estimate. Similar results are found for $\sigma$, where the re-estimation shows a better performance.

\begin{table}[htbp]
	\centering
	\caption{\textbf{Estimation External Parameters}}
	\begin{tabular}{ccccccccc}
		\toprule
		\toprule
		&       & \multicolumn{3}{c}{$\alpha$} &       & \multicolumn{3}{c}{$\sigma$} \\
		\cmidrule{3-5}\cmidrule{7-9}    \textbf{Statistic} &       & $\hat{\alpha}$ &       & $\hat{\hat{\alpha}}$ &       & $\hat{\sigma}$ &       & $\hat{\hat{\sigma}}$ \\
		\cmidrule{1-1}\cmidrule{3-3}\cmidrule{5-5}\cmidrule{7-7}\cmidrule{9-9}
		\textbf{Mean} &       & 41.6725 &       & 23.5121 &       & 1.0697 &       & 1.0975 \\
		\textbf{Median} &       & 41.568 &       & 23.4711 &       & 1.0697 &       & 1.0976 \\
		\textbf{Mode} &       & 33.962 &       & 18.6417 &       & 1.0327 &       & 1.0581 \\
		\textbf{St. Deviation} &       & 2.3421 &       & 1.6915 &       & 0.0119 &       & 0.012 \\
		\bottomrule
		\bottomrule
		\multicolumn{9}{p{11cm}}{\footnotesize{Results obtained for 1000 simulations with 4000 observations for path, with $\alpha=20$, $\sigma=1.1$ and $\gamma=0$.}}
	\end{tabular}
	\label{tab:tabla2.1}
\end{table}

Under the context of the second estimation phase, the Fourier analysis is executed on the 1000 paths of $\hat{\mu}(t)$. With the results obtained by each path, the basic statistics (mean, median, mode) for $\hat{a}_{k}$ and $\hat{\phi}_{k}$ are calculated after the estimates have been ordered, as shown in Table \ref{tab:tabla3.1}. The results indicate that the mean and the median give a good estimate of the amplitude and phase parameters, since the estimates of  $a_{k}$ and $\phi_{k}$ for 1000 simulations are close to the real values with the that was simulated $\mu(t)$. Thus, the median exhibits the best estimators for $a_{k}$ in contrast to the mean, situation that can not be guaranteed with $\phi_{k}$.

\begin{table}[htbp]
	\centering
	\caption{\textbf{Estimate Internal Parameters}}
	\begin{tabular}{ccccccccccccc}
		\toprule
		\toprule
		\multirow{2}[4]{*}{\textbf{Statistic}} &       & \multicolumn{5}{c}{\textbf{Parameters Base}}     &       & \multicolumn{5}{c}{\textbf{Estimation}} \\
		\cmidrule{3-7}\cmidrule{9-13}          &       & $k$     &       & $a_{k}$    &       & $\phi_{k}$    &       & $k$     &       & $a_{k}$    &       & $\phi_{k}$ \\
		\cmidrule{1-1}\cmidrule{3-3}\cmidrule{5-5}\cmidrule{7-7}\cmidrule{9-9}\cmidrule{11-11}\cmidrule{13-13}    \multirow{10}[1]{*}{\textbf{Mean}} &       & 0     &       & 7.3728 &       & 0.0000 &       & 0     &       & 7.3728 &       & 0.0000 \\
		&       & 2     &       & 0.0786 &       & 0.6331 &       & 2     &       & 0.0878 &       & 0.6970 \\
		&       & 4     &       & 0.1664 &       & 2.0853 &       & 4     &       & 0.1642 &       & 1.8211 \\
		&       & 9     &       & 0.1576 &       & -2.1316 &       & 9     &       & 0.1592 &       & -2.1682 \\
		&       & 10    &       & 0.2074 &       & -1.4149 &       & 10    &       & 0.2016 &       & -1.4802 \\
		&       & 12    &       & 0.1376 &       & -1.0862 &       & 12    &       & 0.1354 &       & -1.0678 \\
		&       & 13    &       & 0.1380 &       & 2.6551 &       & 13    &       & 0.1366 &       & 2.4737 \\
		&       & 15    &       & 0.1626 &       & 2.0512 &       & 15    &       & 0.1538 &       & 1.8089 \\
		&       & 16    &       & 0.0964 &       & -1.8092 &       & 16    &       & 0.0948 &       & -1.8894 \\
		&       & 20    &       & 0.1756 &       & -1.8587 &       & 20    &       & 0.1598 &       & -1.9915 \\
		&       &       &       &       &       &       &       &       &       &       &       &  \\
		\multirow{10}[0]{*}{\textbf{Median}} &       & 0     &       & 7.3728 &       & 0.0000 &       & 0     &       & 7.3732 &       & 0.0000 \\
		&       & 2     &       & 0.0786 &       & 0.6331 &       & 2     &       & 0.0832 &       & 0.6212 \\
		&       & 4     &       & 0.1664 &       & 2.0853 &       & 4     &       & 0.1658 &       & 2.0388 \\
		&       & 9     &       & 0.1576 &       & -2.1316 &       & 9     &       & 0.1582 &       & -2.2056 \\
		&       & 10    &       & 0.2074 &       & -1.4149 &       & 10    &       & 0.2036 &       & -1.4966 \\
		&       & 12    &       & 0.1376 &       & -1.0862 &       & 12    &       & 0.1352 &       & -1.1902 \\
		&       & 13    &       & 0.1380 &       & 2.6551 &       & 13    &       & 0.1370 &       & 2.5273 \\
		&       & 15    &       & 0.1626 &       & 2.0512 &       & 15    &       & 0.1552 &       & 1.9063 \\
		&       & 16    &       & 0.0964 &       & -1.8092 &       & 16    &       & 0.0942 &       & -1.9416 \\
		&       & 20    &       & 0.1756 &       & -1.8587 &       & 20    &       & 0.1610 &       & -2.0241 \\
		&       &       &       &       &       &       &       &       &       &       &       &  \\
		\multirow{10}[1]{*}{\textbf{Mode}} &       & 0     &       & 7.3728 &       & 0.0000 &       & 0     &       & 7.3353 &       & 0.0000 \\
		&       & 2     &       & 0.0786 &       & 0.6331 &       & 2     &       & 0.0418 &       & -1.7076 \\
		&       & 4     &       & 0.1664 &       & 2.0853 &       & 4     &       & 0.0404 &       & -2.8641 \\
		&       & 9     &       & 0.1576 &       & -2.1316 &       & 9     &       & 0.0446 &       & -2.6001 \\
		&       & 10    &       & 0.2074 &       & -1.4149 &       & 10    &       & 0.0502 &       & -2.5637 \\
		&       & 12    &       & 0.1376 &       & -1.0862 &       & 12    &       & 0.0820 &       & -1.6728 \\
		&       & 13    &       & 0.138 &       & 2.6551 &       & 13    &       & 0.0526 &       & -2.5862 \\
		&       & 15    &       & 0.1626 &       & 2.0512 &       & 15    &       & 0.057 &       & -2.3391 \\
		&       & 16    &       & 0.0964 &       & -1.8092 &       & 16    &       & 0.0396 &       & -2.8103 \\
		&       & 20    &       & 0.1756 &       & -1.8587 &       & 20    &       & 0.0390 &       & -2.7237 \\
		\bottomrule
		\bottomrule
		\multicolumn{13}{p{12.5cm}}{\footnotesize{Results obtained for 1000 simulations with 4000 observations for path, with a sinusoidal sum of 10 cosines. The parameters base were chosen randomly, with $a_{k}$ for a range between 0.006 and 0.22, the $k$ indicator was randomly chosen between 0 and 20 and the phase angle $\phi_{k}$ was chosen randomly between -2.9 and 2.9. For each path the results are sorted from lowest to highest with reference to $k$.}}
	\end{tabular}
	\label{tab:tabla3.1}
\end{table}

\begin{figure}[htbp]
	\centering
	\includegraphics[width=0.9\linewidth]{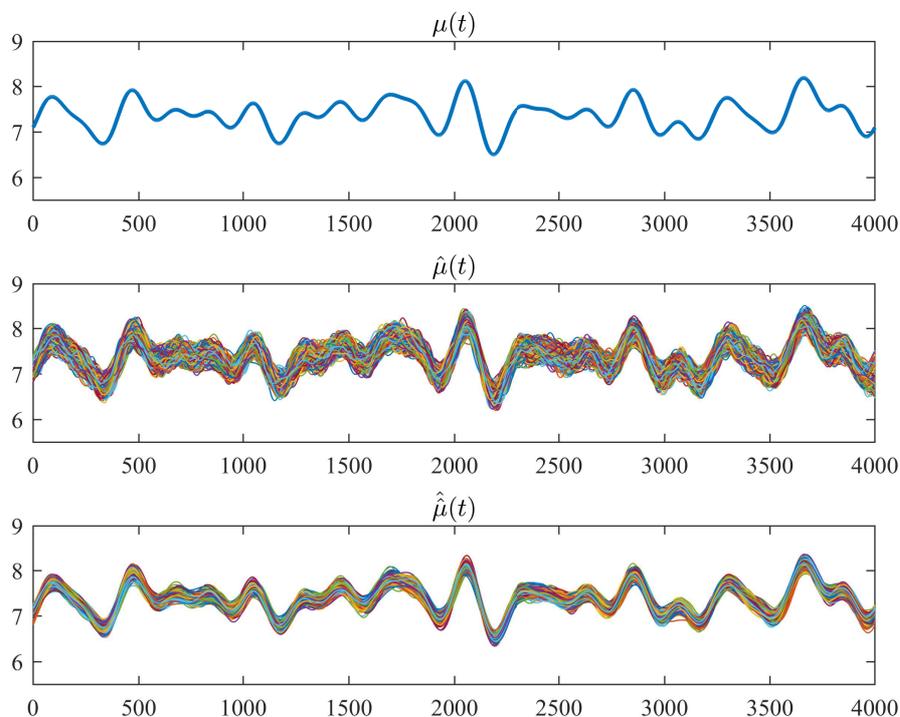}
	\caption{\textbf{Estimates of $\mu(t)$ for all paths}}
	\medskip
	\begin{minipage}{0.7\textwidth} 
		{\footnotesize{1000 paths for  $\hat{\hat{\mu}}(t)$ and $\hat{\mu}(t)$ are calculated for a total of 4000 observations and $\Delta t=\frac{1}{250}$, with the parameters defined in Table \ref{tab:tabla1-1}. The first sub-figure shows the dynamics of $\mu(t)$, the sub-second figure the dynamics for the 1000 paths of $\hat{\mu}(t)$, and the third sub-figure the dynamics of the 1000 paths of $\hat{\hat{\mu}}(t)$.\par}}
	\end{minipage}
	\label{fig:3}
\end{figure}

\begin{figure}[htbp]
	\centering
	\includegraphics[width=0.89\linewidth]{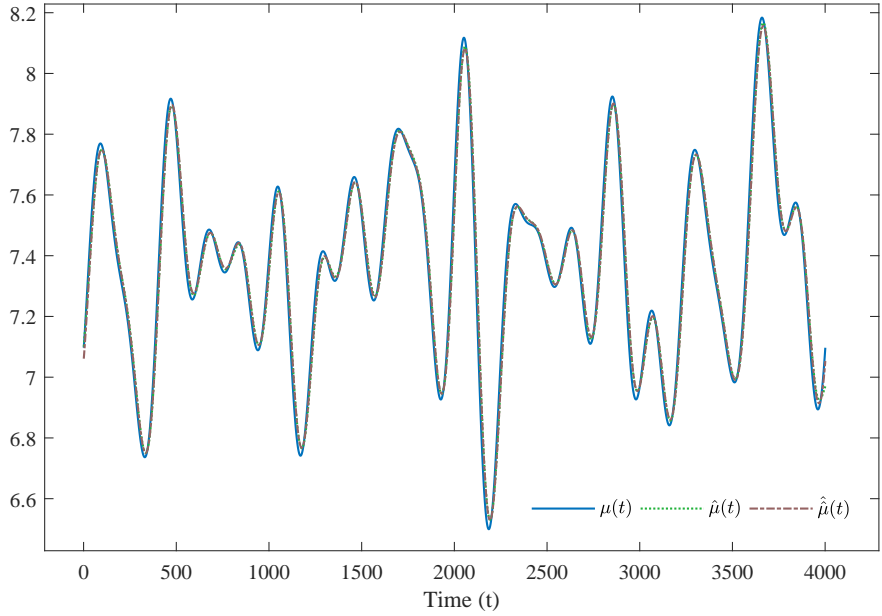}
	\caption{\textbf{Average estimates of $\mu(t)$}}
	\medskip
	\begin{minipage}{0.7\textwidth} 
		{\footnotesize{1000 paths of $\hat{\hat{\mu}}(t)$ and $\hat{\mu}(t)$ are calculated for a total of 4000 observations and a $\Delta t=\frac{1}{250}$, then the average is taken at each point of time for all trajectories of $\hat{\hat{\mu}}(t)$ and $\hat{\mu}(t)$.\par}}
	\end{minipage}
	\label{fig:4}
\end{figure}

With respect to the estimation and re-estimation of $\mu(t)$ the results indicate that both $\hat{\mu}(t)$ and $\hat{\hat{\mu}}(t)$ are good approximations to $\mu(t)$ as shown in the Figure \ref{fig:3} where all the paths of estimation and re-estimation are presented, as well as the dynamics of $\mu(t)$, note that the paths of $\hat{\hat{\mu}}(t)$ are less scattered compared to $\hat{\mu}(t)$. Once we proceed to take the average for each point in the time of the estimation and re-estimation paths, we observe that the behavior of $\mu(t)$, $\hat{\mu}(t)$ and $\hat{\hat{\mu}}(t)$ do not have greater differences as show in Figure \ref{fig:4}. Note that the average estimate of $\hat{\mu}(t)$ and $\hat{\hat{\mu}}(t)$ are very close to original $\mu(t)$, however, in practice there is only one single path composed by discrete observations from which must be extracted all possible information. Hence the importance of having a better approximation.

The table \ref{tab:tabla5.1} shows the RMS for a random path of $\hat{\mu}(t)$ and $\hat{\hat{\mu}}(t)$ with respect to $\mu(t)$ (RMS 1), and the average RMS for each time point of the paths of $\hat{\mu}(t)$ and $\hat{\hat{\mu}}(t)$ with respect to $\mu(t)$ (RMS 2). The results indicate that the RMS error in both cases exhibits a better performance for the re-estimation in contrast with the estimation, in particular in RMS 1 the error in the re-estimation is less than 50\% of the error present in the estimation.  

\begin{table}[htbp]
	\centering
	\caption{\textbf{RMS Estimates}}
	\begin{tabular}{ccccc}
		\toprule
		\toprule
		\textbf{Statistic} &       & $\hat{\mu}(t)$  &       & $\hat{\hat{\mu}}(t)$ \\
		\cmidrule{1-1}\cmidrule{3-3}\cmidrule{5-5}
		\textbf{RMS 1} &       & 0.012423 &       & 0.003873 \\
		\textbf{RMS 2} &       & 0.001665 &       & 0.001645 \\
		\bottomrule
		\bottomrule
		\multicolumn{5}{p{7cm}}{\footnotesize{RMS 1 is calculated from a randomly chosen path for $\hat{\mu}(t)$ and $\hat{\hat{\mu}}(t)$  with respect to $\mu(t)$, while RMS 2 is calculated from the average of all trajectories for each point in time for $\hat{\mu}(t)$ and $\hat{\hat{\mu}}(t)$  with respect to $\mu(t)$.}}
	\end{tabular}
	\label{tab:tabla5.1}
\end{table}

In this way, the previous estimation executed by maximum likelihood provides an accurate estimate of the parameters, where the estimators are consistent, asymptotically unbiased and of minimum variance. Thus, Table \ref{tab:tabla1.2-1} exhibits the estimation of $\alpha$ and $\sigma$ in their first and second estimation phase for different $\Delta t$, the results suggest that with $\Delta t$ smaller and with increasing time periods the average estimates tend to be closer to the real values, also standard deviation is reduced significantly especially for $\alpha$.

Additionally the Fourier analysis allows to obtain an efficient representation of the mean reversion level from a sample of discrete data, with amplitude parameters and phase angles close to the base parameters from which the mean reversion level was simulated. It is important to note that results statistically similar to those found previously are obtained for $\gamma=1$ and $\gamma=\frac{1}{2}$ as shown in appendix \ref{Anexo2.1} and \ref{Anexo3} respectively.

\begin{table}[htbp]
	\centering
	\caption{\textbf{Estimated Parameters for Different $\Delta t$}}
	\begin{tabular}{ccccccccccc}
		\toprule
		\toprule
		&       &       &       & \multicolumn{3}{c}{\textbf{$\alpha$}} &       & \multicolumn{3}{c}{\textbf{$\sigma$}} \\
		\cmidrule{5-7}\cmidrule{9-11}    \textbf{$\Delta t$} &       & \textbf{Statistic} &       & \textbf{$\hat{\alpha}$} &       & \textbf{$\hat{\hat{\alpha}}$} &       & \textbf{$\hat{\sigma}$} &       & \textbf{$\hat{\hat{\sigma}}$} \\
		\cmidrule{1-1}\cmidrule{3-3}\cmidrule{5-5}\cmidrule{7-7}\cmidrule{9-9}\cmidrule{11-11}
		\multirow{2}[0]{*}{$\Delta t_{10}$} &       & \textbf{Mean} &       & 19.9951 &       & 19.9953 &       & 1.1169 &       & 1.0999 \\
		&       & \textbf{Std} &       & 0.0078 &       & 0.008 &       & 0.0504 &       & 0.013 \\
		&       &       &       &       &       &       &       &       &       &  \\
		\multirow{2}[0]{*}{$\Delta t_{50}$} &       & \textbf{Mean} &       & 22.7679 &       & 20.469 &       & 1.0802 &       & 1.0965 \\
		&       & \textbf{Std} &       & 0.6554 &       & 0.6225 &       & 0.0126 &       & 0.0127 \\
		&       &       &       &       &       &       &       &       &       &  \\
		\multirow{2}[0]{*}{$\Delta t_{100}$} &       & \textbf{Mean} &       & 26.8759 &       & 20.9736 &       & 1.0778 &       & 1.0976 \\
		&       & \textbf{Std} &       & 1.0967 &       & 0.9716 &       & 0.0121 &       & 0.012 \\
		\bottomrule
		\bottomrule
		\multicolumn{11}{p{12cm}}{\footnotesize{The results are obtained from 1000 simulations with a total of 4000 observations for each $\Delta t$. Thus, $\Delta t_{10}=\frac{1}{10}$, $\Delta t_{50}=\frac{1}{50}$, $\Delta t_{100}=\frac{1}{100}$.}}
	\end{tabular}
	\label{tab:tabla1.2-1}
\end{table}

\section{Conclusions and Comments }\label{conclusions.1}

This paper shows how starting from the discrete observations of a path of a stochastic process with periodic functional tendency, is possible to find the extern's parameters that define the dynamics of this process and the the internal parameters of periodic functional tendency, for later, to generate projections of such process.

In this sense, two phases of estimation are defined, in the first one we obtain estimates of the external parameters that define the process from closed formulas found from the discretized maximum likelihood function, thus, in the second phase, a re-estimation of parameters is executed considering the analysis of Fourier and the method established in the first phase. The results indicate that the proposed estimation method gives estimated parameters close to the real parameters of a mean reversion stochastic process with periodic functional tendency.

This proposed estimation technique has the advantage of providing closed formulas that only depend on the discrete sampled historical data, where the computation time is negligible and its implementation in terms of programming is easy, additionally can be used and implemented in Fourier series with different periods.

\appendix

\section{Appendices}

\subsection{Existence and uniqueness}\label{Anexo1.1}

To guarantee the existence and uniqueness of the solution of the equation (\ref{ede.1}) when $\gamma=0$ and $\gamma=1$, it will be verified that the Lipschitz and linear growth conditions are satisfied for the One-dimensional case, with $t\in\left\{0,T\right\}$, see \cite{Oksendal2013} for more details.

For $\gamma=0$, 

\begin{equation}
\label{uniexis1.1}
\left|\alpha(\mu(t)-x)\right|+\left|\sigma\right|=\alpha\left|\mu(t)-x\right|+\sigma\le \alpha\left|\mu(t)\right|+\alpha\left|x\right|+\sigma
\end{equation}

On the other hand,

\begin{equation*}
-1\le \cos(2\pi tk+\phi_{k}) \le 1; \hspace{0.5cm} \forall_{t}\in[0,T] \hspace{0.3cm} y \hspace{0.3cm} k\in\mathbb{Z}^{+} 
\end{equation*}

i.e.,

\begin{equation*}
\left|\cos(2\pi tk+\phi_{k})\right|\le 1; \hspace{0.5cm} \left|a_{k}\cos(2\pi tk+\phi_{k})\right|\le a_{k}
\end{equation*}

Thus, 

\begin{equation*}
\left|\mu(t)\right|=\left|\sum _{k=0}^{n}a_{k}\cos(2\pi tk+\phi_{k})\right|\le \sum _{k=0}^{n}\left|a_{k}\cos(2\pi tk+\phi_{k})\right|
\end{equation*}

For hence,

\begin{equation}
\label{uniexis2.1}
\left|\mu(t)\right|\le\sum_{k=0}^{n} a_{k}\hspace{0.5cm} \text{and}\hspace{0.5cm} \alpha\left|\mu(t)\right|\le\alpha\sum_{k=0}^{n} a_{k}
\end{equation}

This is,

\begin{equation}
\label{uniexis3.1}
\alpha\left|\mu(t)\right|+\sigma+\alpha\left|x\right|\le \alpha\sum_{k=0}^{n}a_{k}+\sigma+\alpha\left|x\right|
\end{equation}

Let $C_{*}=\alpha\sum_{k=0}^{n}a_{k}+\sigma$; of (\ref{uniexis1.1}) and (\ref{uniexis3.1}) we get:

\begin{equation*}
\left|\alpha(\mu(t)-x)\right|+\left|\sigma\right| \le C_{*}+\alpha\left|x\right| \le \alpha C_{*}+\alpha C_{*}\left|x\right|	
\hspace{0.15cm} \text{for}\hspace{0.15cm} \alpha>1 \hspace{0.15cm} \text{and}\hspace{0.15cm} C_{*}\ge 1
\end{equation*} 

Finally the linear growth and Lipschitz conditions are respectively guaranteed as follows, 

\begin{equation*}
\left|\alpha(\mu(t)-x)\right|+\left|\sigma\right| \le C_{1}(1+\left|x\right|) \hspace{0.15cm} \text{where} \hspace{0.15cm} C_{1}=\alpha C_{*}
\end{equation*}

On the other hand,

\begin{equation*}
\left|\alpha(\mu(t)-x)-\alpha(\mu(t)-y)\right|+\left|\sigma-\sigma\right|=\alpha\left|-x+y\right|=\alpha\left|x-y\right|\le D_{1}\left|x-y\right|
\end{equation*}

Where, $D_{1}=\frac{\alpha }{\epsilon_{i}}$; $\epsilon_{i}=1,2,3...$.

In this way the existence and uniqueness of the solution of the stochastic differential equation of additive noise is guaranteed.

For $\gamma=1$, 

\begin{equation*}
\left|\alpha(\mu(t)-x)\right|+\left|\sigma x\right|=\alpha\left|\mu(t)-x\right|+\sigma\left|x\right| \le \alpha\left|\mu(t)\right|+\alpha\left|x\right|+\sigma\left|x\right|=\alpha\left|\mu(t)\right|+(\alpha+\sigma)\left|x\right|
\end{equation*}

From the equation (\ref{uniexis2.1}) 

\begin{equation*}
\left|\alpha(\mu(t)-x)\right|+\left|\sigma x\right| \le \alpha\sum_{k=0}^{n}a_{k}+(\alpha+\sigma)\left|x\right| \le \alpha(\alpha+\sigma)\sum_{k=0}^{n}a_{k}+(\alpha+\sigma)\alpha\sum_{k=0}^{n}a_{k}\left|x\right|
\end{equation*}

For $\alpha>1$. 

Finally the linear growth and Lipschitz conditions are respectively guaranteed as follows, 

\begin{equation*}
\left|\alpha\mu(t)-x\right|+\left|\sigma x \right| \le C_{2}(1+\left|x\right|) \hspace{0.5cm} \text{where} \hspace{0.5cm} C_{2}=\alpha(\alpha+\sigma)\sum_{k=0}^{n}a_{k}
\end{equation*}

On the other hand,

\begin{equation*}
\left|\alpha(\mu(t)-x)-\alpha(\mu(t)-y)\right|+\left|\sigma x-\sigma y\right|=\alpha\left|-x+y\right|+\sigma\left|x-y\right|=(\alpha+\sigma)\left|x-y\right| \le D_{2}\left|x-y\right|
\end{equation*}

Where $D_{2}=\frac{\alpha+\sigma}{\epsilon_{i}}$ with $\epsilon_{i}=1,2,3...$.

For the case $\gamma=\frac{1}{2}$ a procedure similar is made. 

\subsection{Estimates for $\gamma=1$}\label{Anexo2.1}

The estimates with $\gamma=1$ are made for 1000 simulations with 4000 observations and $\Delta t=\frac{1}{250}$, additionally $\alpha=30$ and $\sigma=0.2$. Phase 1 and 2 estimation are done as described in the estimation method where $\mu$ is simulated with the phase angles ($\phi_{k}$), amplitude parameters ($a_{k}$) and indicators ($k$) defined in Table \ref{tab:tabla1-1}, and the estimation of $m(t)$ is done with the Hodrick-Prescott filter with a smoothing parameter $\lambda=400000$.

\begin{table}[htbp]
	\centering
	\caption*{\textbf{Estimation External Parameters $\gamma=1$}}
	\begin{tabular}{ccccccccc}
		\toprule
		\toprule
		&       & \multicolumn{3}{c}{$\alpha$} &       & \multicolumn{3}{c}{$\sigma$} \\
		\cmidrule{3-5}\cmidrule{7-9}    \textbf{Statistic} &       & $\hat{\alpha}$ &       & $\hat{\hat{\alpha}}$ &       & $\hat{\sigma}$ &       & $\hat{\hat{\sigma}}$ \\
		\cmidrule{1-1}\cmidrule{3-3}\cmidrule{5-5}\cmidrule{7-7}\cmidrule{9-9}
		\textbf{Mean} &       & 38.1563 &       & 31.6175 &       & 0.1981 &       & 0.1997 \\
		\textbf{Median} &       & 38.0858 &       & 31.6288 &       & 0.1981 &       & 0.1998 \\
		\textbf{Mode} &       & 32.7885 &       & 25.9630 &       & 0.1898 &       & 0.1916 \\
		\textbf{St. Deviation} &       & 1.9723 &       & 1.7880 &       & 0.0022 &       & 0.0022 \\
		\bottomrule
		\bottomrule
		\multicolumn{9}{p{11cm}}{\footnotesize{Results obtained for 1000 simulations with 4000 observations for path, with $\alpha=30$, $\sigma=0.2$ and $\gamma=1$.}}
	\end{tabular}
	\label{tab:tablaA2.1}
\end{table}

\begin{table}[htbp]
	\centering
	\caption*{\textbf{RMS for L Sinusoidal Sums $\gamma=1$}}
	\resizebox{\textwidth}{!}{
	\begin{tabular}{ccccccccccccccc}
		\toprule
		\toprule
		\textbf{L-sum}& & 7  &   & 8 &    & 9 &    & 10&    & 11&    & 12  &  & 13 \\
		\textbf{ RMS }& & 0.001876& & 0.001633& & 0.001735& & 0.001216 & &0.000002 & &0.000001 && 0.000000 \\
		\bottomrule
		\bottomrule
		\multicolumn{15}{p{18cm}}{\footnotesize{1000 paths of $\hat{\hat{\mu}}(t)$ are calculated with Fourier analysis by each sinusoidal sum with cosines 6, 7, 8, 9, 10, 11, 12 and 13. Then, the RMS of $\hat{\hat{\mu}}(t)$ with L sinusoidal sums, with respect to $\hat{\hat{\mu}}(t)$ with L-1 sinusoidal sums is calculated, thus, the RMS of L-sum 7 is calculated as $\frac{\sum (\hat{\hat{\mu}}(t)_{6}-\hat{\hat{\mu}}(t)_{7})^{2}}{n}$ where $\hat{\hat{\mu}}(t)_{6}$ is the average for each point in time of the paths of $\hat{\hat{\mu}}(t)$ with 6 cosines in the sinusoidal sum.}}
	\end{tabular}}
	\label{tab:tabla1.A1.1}
\end{table}

\begin{table}[htbp]
	\centering
	\caption*{\textbf{RMS Estimates $\gamma=1$}}
	\begin{tabular}{ccccc}
		\toprule
		\toprule
		\textbf{Statistic} &       & $\hat{\mu}(t)$  &       & $\hat{\hat{\mu}}(t)$ \\
		\cmidrule{1-1}\cmidrule{3-3}\cmidrule{5-5}
		\textbf{RMS 1} &       & 0.008475 &       & 0.004473 \\
		\textbf{RMS 2} &       & 0.001909 &       & 0.001749 \\
		\bottomrule
		\bottomrule
		\multicolumn{5}{p{7cm}}{\footnotesize{RMS 1 is calculated from a randomly chosen path for $\hat{\mu}(t)$ and $\hat{\hat{\mu}}(t)$  with respect to $\mu(t)$, while RMS 2 is calculated from the average of all paths for each point in time for $\hat{\mu}(t)$ and $\hat{\hat{\mu}}(t)$  with respect to $\mu(t)$.}}
	\end{tabular}
	\label{tab:tablaA5.1}
\end{table}

\begin{table}[htbp]
	\centering
	\caption*{\textbf{Estimate Internal Parameters $\gamma=1$}}
	\begin{tabular}{ccccccccccccc}
		\toprule
		\toprule
		\multirow{2}[4]{*}{\textbf{Statistic}} &       & \multicolumn{5}{c}{\textbf{Parameters Base}}     &       & \multicolumn{5}{c}{\textbf{Estimation}} \\
		\cmidrule{3-7}\cmidrule{9-13}          &       & $k$     &       & $a_{k}$    &       & $\phi_{k}$    &       & $k$     &       & $a_{k}$    &       & $\phi_{k}$ \\
		\cmidrule{1-1}\cmidrule{3-3}\cmidrule{5-5}\cmidrule{7-7}\cmidrule{9-9}\cmidrule{11-11}\cmidrule{13-13}    \multirow{10}[1]{*}{\textbf{Mean}} &       & 0     &       & 7.3728 &       & 0.0000 &       & 0     &       & 7.3723 &       & 0.0000 \\
		&       & 2     &       & 0.0786 &       & 0.6331 &       & 2     &       & 0.0820 &       & 0.6681 \\
		&       & 4     &       & 0.1664 &       & 2.0853 &       & 4     &       & 0.1658 &       & 2.0419 \\
		&       & 9     &       & 0.1576 &       & -2.1316 &       & 9     &       & 0.1566 &       & -2.1451 \\
		&       & 10    &       & 0.2074 &       & -1.4149 &       & 10    &       & 0.2012 &       & -1.4376 \\
		&       & 12    &       & 0.1376 &       & -1.0862 &       & 12    &       & 0.1306 &       & -1.1104 \\
		&       & 13    &       & 0.1380 &       & 2.6551 &       & 13    &       & 0.1300 &       & 2.6246 \\
		&       & 15    &       & 0.1626 &       & 2.0512 &       & 15    &       & 0.1434 &       & 2.0341 \\
		&       & 16    &       & 0.0964 &       & -1.8092 &       & 16    &       & 0.0852 &       & -1.8505 \\
		&       & 20    &       & 0.1756 &       & -1.8587 &       & 20    &       & 0.1260 &       & -1.9020 \\
		&       &       &       &       &       &       &       &       &       &       &       &  \\
		\multirow{10}[0]{*}{\textbf{Median}} &       & 0     &       & 7.3728 &       & 0.0000 &       & 0     &       & 7.3722 &       & 0.0000 \\
		&       & 2     &       & 0.0786 &       & 0.6331 &       & 2     &       & 0.0814 &       & 0.6574 \\
		&       & 4     &       & 0.1664 &       & 2.0853 &       & 4     &       & 0.1664 &       & 2.0811 \\
		&       & 9     &       & 0.1576 &       & -2.1316 &       & 9     &       & 0.1566 &       & -2.1464 \\
		&       & 10    &       & 0.2074 &       & -1.4149 &       & 10    &       & 0.2016 &       & -1.4385 \\
		&       & 12    &       & 0.1376 &       & -1.0862 &       & 12    &       & 0.1298 &       & -1.1235 \\
		&       & 13    &       & 0.1380 &       & 2.6551 &       & 13    &       & 0.1306 &       & 2.6300 \\
		&       & 15    &       & 0.1626 &       & 2.0512 &       & 15    &       & 0.1428 &       & 2.0403 \\
		&       & 16    &       & 0.0964 &       & -1.8092 &       & 16    &       & 0.0856 &       & -1.8534 \\
		&       & 20    &       & 0.1756 &       & -1.8587 &       & 20    &       & 0.1262 &       & -1.9028 \\
		&       &       &       &       &       &       &       &       &       &       &       &  \\
		\multirow{10}[1]{*}{\textbf{Mode}} &       & 0     &       & 7.3728 &       & 0.0000 &       & 0     &       & 7.3325 &       & 0.0000 \\
		&       & 2     &       & 0.0786 &       & 0.6331 &       & 2     &       & 0.0344 &       & -0.3107 \\
		&       & 4     &       & 0.1664 &       & 2.0853 &       & 4     &       & 0.0440 &       & -2.3652 \\
		&       & 9     &       & 0.1576 &       & -2.1316 &       & 9     &       & 0.0998 &       & -2.4985 \\
		&       & 10    &       & 0.2074 &       & -1.4149 &       & 10    &       & 0.0562 &       & -2.5378 \\
		&       & 12    &       & 0.1376 &       & -1.0862 &       & 12    &       & 0.0822 &       & -1.5151 \\
		&       & 13    &       & 0.1380 &       & 2.6551 &       & 13    &       & 0.0488 &       & -2.1121 \\
		&       & 15    &       & 0.1626 &       & 2.0512 &       & 15    &       & 0.0614 &       & -1.8791 \\
		&       & 16    &       & 0.0964 &       & -1.8092 &       & 16    &       & 0.0398 &       & -2.6621 \\
		&       & 20    &       & 0.1756 &       & -1.8587 &       & 20    &       & 0.0914 &       & -2.2005 \\
		\bottomrule
		\bottomrule
		\multicolumn{13}{p{12.5cm}}{\footnotesize{Results obtained for 1000 simulations with 4000 observations for path, with a sinusoidal sum of 10 cosines. The parameters base were chosen randomly, with $a_{k}$ for a range between 0.006 and 0.22, the $k$ indicator was randomly chosen between 0 and 20 and the phase angle $\phi_{k}$ was chosen randomly between -2.9 and 2.9. For each path the results are sorted from lowest to highest with reference to $k$.}}
	\end{tabular}
	\label{tab:tablaA3.1}
\end{table}

\subsection{Estimates for $\gamma=\frac{1}{2}$}\label{Anexo3}

The estimates with $\gamma=\frac{1}{2}$ are made for 1000 simulations with 4000 observations and $\Delta t=\frac{1}{250}$, additionally $\alpha=23$ and $\sigma=0.6$. Phase 1 and 2 estimation are done as described in the estimation method where $\mu$ is simulated with the phase angles ($\phi_{k}$), amplitude parameters ($a_{k}$) and indicators ($k$) defined in Table \ref{tab:tabla1-1}, and the estimation of $m(t)$ is done with the Hodrick-Prescott filter with a smoothing parameter $\lambda=400000$.

\begin{table}[htbp]
	\centering
	\caption*{\textbf{Estimation External Parameters $\gamma=\frac{1}{2}$}}
	\begin{tabular}{ccccccccc}
		\toprule
		\toprule
		&       & \multicolumn{3}{c}{$\alpha$} &       & \multicolumn{3}{c}{$\sigma$} \\
		\cmidrule{3-5}\cmidrule{7-9}    \textbf{Statistic} &       & $\hat{\alpha}$ &       & $\hat{\hat{\alpha}}$ &       & $\hat{\sigma}$ &       & $\hat{\hat{\sigma}}$ \\
		\cmidrule{1-1}\cmidrule{3-3}\cmidrule{5-5}\cmidrule{7-7}\cmidrule{9-9}
		\textbf{Mean} &       & 32.6656 &       & 25.3054 &       & 0.5929 &       & 0.5985 \\
		\textbf{Median} &       & 32.6135 &       & 25.2131 &       & 0.5930 &       & 0.5986 \\
		\textbf{Mode} &       & 27.8607 &       & 20.5132 &       & 0.5678 &       & 0.5741 \\
		\textbf{St. Deviation} &       & 1.8896 &       & 1.6816  &       & 0.0065 &       & 0.0066 \\
		\bottomrule
		\bottomrule
		\multicolumn{9}{p{11cm}}{\footnotesize{Results obtained for 1000 simulations with 4000 observations for path, with $\alpha=23$, $\sigma=0.6$ and $\gamma=\frac{1}{2}$.}}
	\end{tabular}
	\label{tab:tabla2.2}
\end{table}

\begin{table}[htbp]
	\centering
	\caption*{\textbf{RMS for L Sinusoidal Sums $\gamma=\frac{1}{2}$}}
	\resizebox{\textwidth}{!}{
	\begin{tabular}{ccccccccccccccc}
		\toprule
		\toprule
		\textbf{L-sum} && 7 &    & 8 &    & 9 &    & 10 &   & 11&    & 12 &   & 13 \\
		\textbf{ RMS }& & 0.001383& & 0.001099& & 0.001016& & 0.000644 & &0.000022& & 0.000003& & 0.000001 \\
		\bottomrule
		\bottomrule
		\multicolumn{15}{p{18cm}}{\footnotesize{1000 paths of $\hat{\hat{\mu}}(t)$ are calculated with Fourier analysis by each sinusoidal sum with cosines 6, 7, 8, 9, 10, 11, 12 and 13. Then, the RMS of $\hat{\hat{\mu}}(t)$ with L sinusoidal sums, with respect to $\hat{\hat{\mu}}(t)$ with L-1 sinusoidal sums is calculated, thus, the RMS of L-sum 7 is calculated as $\frac{\sum (\hat{\hat{\mu}}(t)_{6}-\hat{\hat{\mu}}(t)_{7})^{2}}{n}$ where $\hat{\hat{\mu}}(t)_{6}$ is the average for each point in time of the paths of $\hat{\hat{\mu}}(t)$ with 6 cosines in the sinusoidal sum.}}
	\end{tabular}}
	\label{tab:tabla1.A.1}
\end{table}

\begin{table}[htbp]
	\centering
	\caption*{\textbf{RMS Estimates $\gamma=\frac{1}{2}$}}
	\begin{tabular}{ccccc}
		\toprule
		\toprule
		\textbf{Statistic} &       & $\hat{\mu}(t)$  &       & $\hat{\hat{\mu}}(t)$ \\
		\cmidrule{1-1}\cmidrule{3-3}\cmidrule{5-5} 
		\textbf{RMS 1} &       & 0.0165092 &       & 0.0075371 \\
		\textbf{RMS 2} &       & 0.0022999 &       & 0.0022304 \\
		\bottomrule
		\bottomrule
		\multicolumn{5}{p{7cm}}{\footnotesize{RMS 1 is calculated from a randomly chosen path for $\hat{\mu}(t)$ and $\hat{\hat{\mu}}(t)$  with respect to $\mu(t)$, while RMS 2 is calculated from the average of all paths for each point in time for $\hat{\mu}(t)$ and $\hat{\hat{\mu}}(t)$  with respect to $\mu(t)$.}}
	\end{tabular}
	\label{tab:tabla5A.1}
\end{table}

\begin{table}[htbp]
	\centering
	\caption*{\textbf{Estimate Internal Parameters $\gamma=\frac{1}{2}$}}
	\begin{tabular}{ccccccccccccc}
		\toprule
		\toprule
		\multirow{2}[4]{*}{\textbf{Statistic}} &       & \multicolumn{5}{c}{\textbf{Parameters Base}}     &       & \multicolumn{5}{c}{\textbf{Estimation}} \\
		\cmidrule{3-7}\cmidrule{9-13}          &       & $k$     &       & $a_{k}$    &       & $\phi_{k}$    &       & $k$     &       & $a_{k}$    &       & $\phi_{k}$ \\
		\cmidrule{1-1}\cmidrule{3-3}\cmidrule{5-5}\cmidrule{7-7}\cmidrule{9-9}\cmidrule{11-11}\cmidrule{13-13}    \multirow{10}[1]{*}{\textbf{Mean}} &       & 0     &       & 7.3728 &       & 0.0000 &       & 0     &       & 7.3735 &       & 0.0000 \\
		&       & 2     &       & 0.0786 &       & 0.6331 &       & 2     &       & 0.0992 &       & 0.7879 \\
		&       & 4     &       & 0.1664 &       & 2.0853 &       & 4     &       & 0.1550 &       & 1.6093 \\
		&       & 9     &       & 0.1576 &       & -2.1316 &       & 9     &       & 0.1574 &       & -1.8970 \\
		&       & 10    &       & 0.2074 &       & -1.4149 &       & 10    &       & 0.1930 &       & -1.4631 \\
		&       & 12    &       & 0.1376 &       & -1.0862 &       & 12    &       & 0.1344 &       & -1.0284 \\
		&       & 13    &       & 0.1380 &       & 2.6551 &       & 13    &       & 0.1292 &       & 2.2265 \\
		&       & 15    &       & 0.1626 &       & 2.0512 &       & 15    &       & 0.1412 &       & 1.9505 \\
		&       & 16    &       & 0.0964 &       & -1.8092 &       & 16    &       & 0.0936 &       & -1.4885 \\
		&       & 20    &       & 0.1756 &       & -1.8587 &       & 20    &       & 0.1248 &       & -1.9415 \\
		&       &       &       &       &       &       &       &       &       &       &       &  \\
		\multirow{10}[0]{*}{\textbf{Median}} &       & 0     &       & 7.3728 &       & 0.0000 &       & 0     &       & 7.3737 &       & 0.0000 \\
		&       & 2     &       & 0.0786 &       & 0.6331 &       & 2     &       & 0.0916 &       & 0.6793 \\
		&       & 4     &       & 0.1664 &       & 2.0853 &       & 4     &       & 0.1624 &       & 2.0323 \\
		&       & 9     &       & 0.1576 &       & -2.1316 &       & 9     &       & 0.1580 &       & -2.1401 \\
		&       & 10    &       & 0.2074 &       & -1.4149 &       & 10    &       & 0.1978 &       & -1.4626 \\
		&       & 12    &       & 0.1376 &       & -1.0862 &       & 12    &       & 0.1320 &       & -1.1530 \\
		&       & 13    &       & 0.1380 &       & 2.6551 &       & 13    &       & 0.1298 &       & 2.5735 \\
		&       & 15    &       & 0.1626 &       & 2.0512 &       & 15    &       & 0.1418 &       & 2.0161 \\
		&       & 16    &       & 0.0964 &       & -1.8092 &       & 16    &       & 0.0906 &       & -1.8483 \\
		&       & 20    &       & 0.1756 &       & -1.8587 &       & 20    &       & 0.1250 &       & -1.9390 \\
		&       &       &       &       &       &       &       &       &       &       &       &  \\
		\multirow{10}[1]{*}{\textbf{Mode}} &       & 0     &       & 7.3728 &       & 0.0000 &       & 0     &       & 7.3187 &       & 0.0000 \\
		&       & 2     &       & 0.0786 &       & 0.6331 &       & 2     &       & 0.0404 &       & -2.8082 \\
		&       & 4     &       & 0.1664 &       & 2.0853 &       & 4     &       & 0.0420 &       & -3.0263 \\
		&       & 9     &       & 0.1576 &       & -2.1316 &       & 9     &       & 0.0528 &       & -3.1413 \\
		&       & 10    &       & 0.2074 &       & -1.4149 &       & 10    &       & 0.0488 &       & -2.9586 \\
		&       & 12    &       & 0.1376 &       & -1.0862 &       & 12    &       & 0.0562 &       & -3.1305 \\
		&       & 13    &       & 0.1380 &       & 2.6551 &       & 13    &       & 0.0442 &       & -3.1298 \\
		&       & 15    &       & 0.1626 &       & 2.0512 &       & 15    &       & 0.0504 &       & -2.7634 \\
		&       & 16    &       & 0.0964 &       & -1.8092 &       & 16    &       & 0.0408 &       & -2.9482 \\
		&       & 20    &       & 0.1756 &       & -1.8587 &       & 20    &       & 0.0400 &       & -3.0440 \\
		\bottomrule
		\bottomrule
		\multicolumn{13}{p{12.5cm}}{\footnotesize{Results obtained for 1000 simulations with 4000 observations for path, with a sinusoidal sum of 10 cosines. The parameters base were chosen randomly, with $a_{k}$ for a range between 0.006 and 0.22, the $k$ indicator was randomly chosen between 0 and 20 and the phase angle $\phi_{k}$ was chosen randomly between -2.9 and 2.9. For each path the results are sorted from lowest to highest with reference to $k$.}}
	\end{tabular}
	\label{tab:tabla3A.1}
\end{table}

\bibliography{library}
\bibliographystyle{IEEEtran}

\end{document}